\documentstyle[aps,prl,twocolumn]{revtex}
\input amssym.def
\input amssym
\font\script=eusm10
\font\bbold=msbm10

\newcommand{\SH}{\mbox{\script H}}

\newcommand{\BR}{\mbox{\bbold R}}
\begin{document}
\title{A Lorentz-invariant look at quantum clock synchronization
protocols based on distributed entanglement}
\author{Ulvi Yurtsever and Jonathan P.\ Dowling}
\address{Jet Propulsion Laboratory, California Institute of
Technology, 4800 Oak Grove Drive, Pasadena, California 91109-8099.}
\date{Received: \today}
\maketitle
\begin{abstract}
Recent work has raised the possibility that quantum information theory
techniques can be used to synchronize atomic clocks nonlocally. One of
the proposed algorithms for quantum clock synchronization (QCS) requires
distribution of entangled pure singlets to the synchronizing parties. Such
remote entanglement distribution normally creates a relative phase
error in the distributed singlet state which then
needs to be purified asynchronously. We present a fully relativistic
analysis of the QCS protocol which shows that asynchronous entanglement
purification is not possible, and, therefore, that the proposed QCS scheme
remains incomplete. We discuss possible directions of research
in quantum information theory which may lead to a complete, working QCS
protocol.

~~~~~

\end{abstract}
\parskip 4pt

~~~~~~

{\bf \noindent 1. Clock synchronization with shared singlets}

Suppose a supply of identical
but distinguishable two-state systems (e.g.\ atoms) are available
whose between-state transitions can be manipulated (e.g.\ by
laser pulses). Let $|1 \rangle$ and $|0 \rangle$ denote,
respectively, the excited and
ground states (which span
the internal Hilbert space ${\SH}$) of the prototype two-state system,
and let the energy difference between the
two states be $\Omega$ (we will use units in which $\hbar = c = 1$
throughout this letter). Without loss of generality, we can assume
\begin{equation}
H_0 \, | 0 \rangle = 0 \; , \; \; \; \; \; \; \;
H_0 \, | 1 \rangle = \Omega \, | 1 \rangle \; ,
\end{equation}
where $H_0$ denotes the internal Hamiltonian.
Suppose pairs of these two-state systems are distributed to
two spatially-separated observers Alice and Bob. The Hilbert space of
each pair can be written as ${\SH}_A \otimes {\SH}_B$, where $\otimes$
denotes tensor product between vector spaces.
A (``pure") singlet is the specific entangled quantum state in this product
Hilbert space given by
\begin{equation}
\Psi = \frac{1}{\sqrt{2}}\, (\, |0 \rangle_A \otimes |1 \rangle_B -
| 1 \rangle_A \otimes | 0 \rangle_B ) \;
\end{equation}
[in what follows,
we will omit tensor-product signs in
expressions of the kind Eq.\,(2) unless required for clarity].
Two important properties of the singlet state $\Psi$ are: (i) it is a
``dark" state (invariant up to a multiplicative phase factor)
under the time evolution $U_t \equiv \exp (i t H_0 )$, i.e.\ $(U_t \otimes U_t )\Psi
= e^{i \phi} \Psi$ where $e^{i \phi}$ is an overall phase,
and (ii) it is similarly invariant under all
unitary transformations of the form $U \otimes U$, where $U$ is any
arbitrary unitary map on $\SH$ (not necessarily equal to $U_t$).
Both properties are
needed for the Quantum Clock Synchronization (QCS) protocol of Jozsa {\it et.\
al.} \cite{QCS}, which assumes a supply of such
pure singlet states shared as a resource
between the synchronizing parties Alice and Bob. Specifically,
consider the unitary (Hadamard)
transformation ($\pi /2$ -- pulse followed by the spin operator $\sigma_z$)
on $\SH$ given by
\begin{eqnarray}
| 0 \rangle & \longmapsto & | + \rangle \equiv \frac{1}{\sqrt{2}}\,
(|0\rangle + |1 \rangle) \; , \nonumber \\
| 1 \rangle & \longmapsto & | - \rangle \equiv \frac{1}{\sqrt{2}}\,
(|0\rangle - |1 \rangle) \; .
\end{eqnarray}
Unlike the states $|0\rangle$ and $|1\rangle$, which are dark under time
evolution (they only pick up an overall phase under $U_t$), the states
$|+\rangle$ and $|-\rangle$ are ``clock states" (in other words, they
accumulate an observable relative phase under $U_t$) because of the energy
difference $\Omega$ as specified in Eq.\,(1). Such states can be used
to ``drive" precision clocks in the following way: Start, for example,
with an ensemble of atoms in
the state $|+\rangle$ produced by an initial Hadamard
pulse at time $t_0$, and apply a second
Hadamard pulse at a later time $t_0 +T$.
This leads to a final state at $t_0 + T$
equivalent, up to an overall phase factor, to the state
\begin{equation}
\cos \left( \frac{\Omega}{2} T \right) |0\rangle + i \sin
\left( \frac{\Omega}{2} T \right) |1\rangle \; .
\end{equation}
Measurement of the statistics (relative populations) of ground vs.\
excited atoms in the state Eq.\,(4) then yields a precision measurement
of the time interval $T$; hence clock funcionality
for $|+\rangle$. [In practice (e.g.\ in ``real-world" atomic
clocks), such measurements are used to stabilize the frequency of a
relatively noisy local oscillator (typically a maser), whose
(stabilized) oscillations then drive the ultimate clock readout.]
Now, the invariance of the pure singlet $\Psi$ [Eq.\,(2)] under the Hadamard
transformation Eq.\,(3) can be seen explicitly in the alternative
representation
\begin{equation}
\Psi = \frac{1}{\sqrt{2}}\, (\, |- \rangle_A \otimes |+ \rangle_B -
| + \rangle_A \otimes | - \rangle_B ) \; ,
\end{equation}
and here, in Eq.\,(5), we have the crux of the QCS algorithm of Ref.\,[1]:
The dark, invariant state $\Psi$ shared between Alice and Bob contains
two clock states, one for each observer,
entangled in such a way as to freeze their time evolution. As soon as
Bob (or Alice) performs a measurement on $\Psi$ in the basis $\{
|+\rangle, |-\rangle \}$, thereby destroying the entanglement,
he (or she) starts these two dormant clocks
``simultaneously" in both reference frames (classical communications are
then necessary to sort out which party has the $|+\rangle$ clock and
which party has $|-\rangle$). When used to stabilize identical quantum clocks
at each party's location, these correlated clock states then
provide precise time synchrony between Bob and Alice\cite{Ftnt1}.

It is important to emphasize that the ``energetic" nature of the singlet
state $\Psi$ [Eq.\,(2)] is crucial for the QCS protocol to work. This
is in complete contrast with other quantum information-theory
protocols (such as teleportation\cite{Telep}, quantum cryptographic key
distribution\cite{Crypt}, and others) all of which will work equally well with
degenerate ($\Omega =0$) singlets.

{\bf \noindent 2. ``Impure" singlets, QCS algorithm, and teleportation}

In principle, the QCS protocol as outlined above is rigorously correct
and self contained. If our Universe somehow possessed
primordial energetic singlet states $\Psi$
(left as ``relics" from the Big Bang), the protocol
just described would be
perfectly sufficient to implement ultra-precise clock synchronization
between comoving distant observers.
In practice, however, the QCS algorithm can reasonably be viewed as
simply reducing the problem of clock synchronization to the problem of
distributing pure entanglement to spatially separated
regions of spacetime. To see that the latter is a non-trivial problem,
consider the simplest way one would attempt to distribute entanglement
to remote regions: start with pairs of two-level systems (atoms)
in locally-created pure singlet states $\Psi$ in the form
Eq.\,(2), and transport the two
subsytems separately to the locations of Bob and Alice.
The internal Hamiltonians of the two subsystems
while in transport can be written in the form
\begin{equation}
H_A = H_0 + {H_A}^{\rm ext} \; , \; \; \; \; \; \; \;
H_B = H_0 + {H_B}^{\rm ext} \; ,
\end{equation}
where ${H_A}^{\rm ext}$ and ${H_B}^{\rm ext}$ denote interaction
Hamiltonians arising from the coupling of each subsystem to its
environment, and, unless the
environment that each subsytem is subject to during transport is
precisely controlled, ${H_A}^{\rm ext} \neq {H_B}^{\rm ext}$ in general,
leading to a relative phase offset in the final entangled state.
Furthermore, unless the worldlines of the transported subsytems are
arranged to have precisely the same Lorentz length (proper time),
a further contribution to this phase offset would occur
due to the proper-time delay between the two worldlines
(see the discussion in Sect.\,3 below). The end result is an impure
singlet state
\begin{equation}
\Psi_\delta = \frac{1}{\sqrt{2}}\, (\, |0 \rangle_A |1 \rangle_B -
e^{i \delta} \, | 1 \rangle_A  | 0 \rangle_B ) \; ,
\end{equation}
where $\delta$ is a real phase offset which is fixed but entirely unknown.
[In general, coupling to the environment will lead to other errors such
as bit flips and decoherence, resulting in a mixed state at the end of
the transport process. These kinds of errors, however, are correctable
(after restoring energy degeneracy to the qubit basis
$\{ |0\rangle, \, |1\rangle\}$ if necessary) by using
standard entanglement purification techniques\cite{Bent3}. The phase
error in Eq.\,(7), however, is inextricably mixed with the
synchronization offset between Alice and Bob as we will argue below, and
it cannot be purified asynchronously.]

Although $\Psi_\delta$ is still a dark state under time evolution,
it no longer has the key property of invariance under arbitrary unitary
transformations $U \otimes U$. In particular, a ``magic" equivalent
form like Eq.\,(5) in terms of entangled clock states is not available
for $\Psi_\delta$\cite{Ftnt2}. Instead,
\begin{eqnarray}
\Psi_\delta & = & \left( \frac{1+e^{i\delta}}{2 \sqrt{2}} \right)
(|-\rangle_A | + \rangle_B - |+\rangle_A | - \rangle_B ) \nonumber \\
& + &
\left( \frac{1-e^{i\delta}}{2 \sqrt{2}} \right)
(|+\rangle_A | + \rangle_B - |-\rangle_A | - \rangle_B ) \; ,
\end{eqnarray}
and a measurement by Bob or Alice in the $\{ | +\rangle, \, |-\rangle
\}$ basis will leave the other party's clock in a superposition
of clock states $|+\rangle$ and $|-\rangle$, which, if Bob and Alice
were to follow the above QCS protocol blindly, effectively
introduces an (unknown) synchronization offset of $- \delta / \Omega$
between them.

This connection between $\delta$ and the time synchronization offset
is much easier to understand by adopting a different point of view on
the QCS protocol: one which is based on teleportation\cite{Telep}. Accordingly,
the essence of the  QCS protocol can be viewed as the
teleportation of clock states between Bob and Alice using the singlet
states $\Psi$ (or, in the present case, the impure singlets $\Psi_\delta$).
More explicitly,
suppose Bob and Alice arrange, through prior classical communications,
the teleportation of a known quantum state
$\alpha |0\rangle_{B'} + \beta | 1 \rangle_{B'}$ $\in \SH_{B'}$ from Bob
to Alice via the singlet $\Psi_\delta$. Since the teleported state, as well as
Bob's Bell-basis states\cite{Telep}
\begin{eqnarray}
\Psi^{\pm} & \equiv & \frac{1}{\sqrt{2}} \, (\, |0\rangle_B |
1 \rangle_{B'} \pm |1\rangle_B |0\rangle_{B'} ) \; , \nonumber \\
\Phi^{\pm} & \equiv & \frac{1}{\sqrt{2}} \, (\, |0\rangle_B |
0 \rangle_{B'} \pm |1\rangle_B |1\rangle_{B'} )  \; 
\end{eqnarray}
are in general time dependent, the standard teleportation protocol needs
to be slightly modified in the following way: The parties need to agree
on a time, which we may take without
loss of generality to be $t_B = 0$ as measured
by Bob's local clock, at which the following three actions
will be performed instantaneously by Bob:
(i) prepare an ancillary two-state system $B'$ in the known quantum state
$\alpha |0\rangle_{B'} + \beta | 1 \rangle_{B'}$,
where $\alpha$ and $\beta$ are complex numbers previously agreed on by
the two parties, (ii) select a specific singlet
$\Psi_\delta$ as in Eq.\,(7),
and construct a Bell basis for $\SH_B \otimes \SH_{B'}$
that has the form Eq.\,(9)
at $t_B = 0$, and (iii) perform a measurement in this basis and
communicate its outcome to Alice through a classical channel.
Upon receipt of this outcome, Alice is then to rotate the
(collapsed) quantum state of her half of the
singlet $\Psi_\delta$ (now a vector in the Hilbert space $\SH_A$) by one of
the four unitary operators
\begin{eqnarray}
M_{\Psi^{\pm}} & = & 
{\pm} \, |0\rangle_A \langle 0|_A \; - \;
e^{-i H_0 \, t_A} \, |1 \rangle_A \langle 1|_A \; \; , \nonumber \\
M_{\Phi^{\pm}} & = &
-e^{-i H_0 \, t_A} \, |0\rangle_A \langle 1 |_A
\; \, {\pm} \, \; | 1 \rangle_A \langle 0 |_A \; \; 
\end{eqnarray}
depending on whether the transmitted outcome of Bob's measurement is
one of $\Psi^{+}, \; \Psi^- , \; \Phi^+ $ or $\Phi^-$.
Here $t_A$ denotes Alice's proper time (as measured by her
local clock) at the moment she performs her unitary rotation.
Now let the (unknown) synchronization offset between Bob and Alice be
$\tau$, so that $t_B = t_A + \tau$. It is easy to show that the
state teleported to Alice under this arrangement will have the form
\begin{equation}
\alpha \, |0 \rangle_A \; + \; e^{i (-\Omega \tau + \delta)} \, \beta \,
|1\rangle_A
\end{equation}
as obtained by Alice immediately following her unitary operation on
$\SH_A$.

A number of key results can now be easily read out from Eq.\,(11):
(1) If $\delta =0$, i.e.\ under the same assumption as in the original QCS
protocol\cite{QCS} that the shared singlet states are pure,
the time-synchronization offset $\tau$ can be immediately
determined by Alice (recall that $\alpha$ and $\beta$ are known to
both parties). Hence, the synchronization result of the
QCS protocol can equivalently be achieved through teleportation.
(2) Conversely, if $\tau =0$, i.e.\ if Bob and Alice have their
clocks synchronized
to begin with, {\em or} if $\Omega =0$, i.e.\ if the qubits
spanning the local Hilbert spaces $\SH_A$ and $\SH_B$ are degenerate,
then $\delta$ can be immediately determined by Alice. Hence,
purification of the phase-offset singlet $\Psi_\delta$ is possible under
either of these two conditions. (3) If, on the other hand, none of the
quantities $\Omega, \; \tau$, and $\delta$ vanish, then the two unknowns
$\tau$ and $\delta$ are inextricably mixed in the only phase observable
$- \Omega \tau + \delta$, and asynchronous purification is impossible.

This last conclusion can be greatly clarified and strengthened by a
fully Lorentz-invariant formulation of the above teleportation protocol
(which, as we just argued, is equivalent to the original QCS), and it is
this formulation we will turn to next.

{\bf \noindent 3. Lorentz-invariant analysis of QCS}

The key ingredient in any relativistic discussion of quantum
information theory is the spacetime dependence of the qubit states.
The ``true" Hilbert space to which the quantum state of a singlet belongs is,
accordingly, $L^2(\BR^4 )\otimes \SH_A \otimes L^2(\BR^4 )
\otimes \SH_B$, where each $L^2( \BR^4 )$ is supposed to account for the
spacetime wave function of each two-state system in the entangled pair.
We will assume a flat, Minkowski spacetime
background in what follows, and pretend that the spacetime dependence of
each system's wave function can be approximated
by that of a plane wave. In a more careful treatment, plane waves should
be replaced by localized, normalizable wave packets [which have the
admirable property, unlike plane waves, of being truly in $L^2(\BR^4 )$].

Denote the four velocities of Alice and Bob by $u_A$ and $u_B$,
respectively, so that $u_A \cdot u_A = u_B \cdot u_B = -1$ [we will adopt
the sign convention in which Minkowski metric on $\BR^4$ has the
form $\eta = -dt \otimes dt + dx \otimes dx + dy \otimes dy
+ dz \otimes dz$, and use the
abbreviation $a \cdot b$ to denote $\eta (a,b)$
for any two four-vectors $a$ and $b$]. The wave four-vectors
of Alice's and Bob's atoms then have the form
\begin{equation}
{k^0}_J = m_0 u_J \; \; , \; \; \; \; \; \;
{k^1}_J = (m_0 + \Omega ) u_J \; ,
\end{equation}
where $m_0$ is the ground-state
rest mass of each (identical) two-level atom, and
${k^0}_J$ and ${k^1}_J$ denote the wave vectors corresponding to the ground
and excited states of the atoms, respectively, where $J = A, \;B$. The
plane-wave spacetime dependence of the
wave functions corresponding to the ground and excited states
of each of the atoms can then be written in the form
\begin{equation}
|0\rangle_J \longrightarrow e^{i {k^0}_J \cdot x } \; |0\rangle_J
\; , \; \; \; \; \;
|1\rangle_J \longrightarrow e^{i {k^1}_J \cdot x } \; |1\rangle_J \; ,
\end{equation}
where $J= A, \; B$, and $x$ denotes an arbitrary point (event) in
spacetime (a four-vector). Simple algebra then shows that, up to an overall
phase factor which can always be ignored, the wave function
corresponding to the singlet state Eq.\,(7) can be expressed as a
two-point spacetime function of the form
\begin{equation}
\Psi_\delta (x_1 , x_2 ) = |0\rangle_A | 1 \rangle_B - 
e^{i \Phi_\delta (x_1 , x_2 )} |1\rangle_A |0\rangle_B \; ,
\end{equation}
where $x_1$ and $x_2$ denote spacetime points along the worldlines
of Alice and Bob, respectively, and
$\Phi_\delta (x_1 , x_2 )$ is the
Lorentz-invariant two-point phase function
\begin{equation}
\Phi_\delta (x_1 , x_2 ) \equiv \Omega \, (u_A \cdot x_1 - u_B \cdot x_2 )
+ \delta \; .
\end{equation}
In the important special case where $u_A = u_B = u$, i.e.\ when Alice
and Bob are comoving (and it makes sense
to synchronize their clocks), $\Phi_\delta$ takes the simpler form
\begin{equation}
\Phi_\delta (x_1 , x_2 ) = \Omega \, u \cdot (x_1 -  x_2 )
+ \delta \; .
\end{equation}
In the comoving case Eq.\,(16) (when $u_A = u_B$),
the singlet wave function $\Psi_\delta ( x_1 , x_2 )$ is invariant under
arbitrary Lorentz transformations including translations. This
is in contrast with the general case,
where the phase function $\Phi_\delta (x_1 , x_2 )$ [Eq.\,(15)] does
not have translation invariance. This dependence on the choice of
origin of coordinates is a manifestation of the fact
that $\Psi_\delta$ is not a dark state unless $u_A = u_B$.

The teleportation protocol of the previous section
(which is equivalent to the original QCS protocol of\cite{QCS})
demonstrates that as
long as $x_1$ and $x_2$ are timelike separated events in spacetime,
the relative phase $\Phi_\delta (x_1 , x_2 )$ can
be directly observed by Alice and Bob via
quantum measurements followed by classical communication of the outcomes.
Conversely, since the wave function contains all
knowledge that can ever be obtained about a quantum system,
the {\em only} observable associated with the singlet
state $\Psi_\delta$ which contains any information about $\delta$
is $\Phi_\delta (x_1 , x_2 )$.
Focusing now on the comoving case $u_A = u_B$,
this implies that the phase offset
$\delta$ is {\em not} observable in isolation; only the
combination two-point function $\delta + \Omega \, u \cdot (x_1 -  x_2 )$
[Eq.\,(16)] is accessible to direct measurement. On the other hand,
clock synchronization between Bob and Alice is
equivalent to identification of pairs of events $(x^{(i)}_1 , x^{(i)}_2 )$
such that $u \cdot (x^{(i)}_1 - x^{(i)}_2 ) =0 $. Therefore,
by making a sequence of measurements of the relative
phase function $\Phi_\delta (x_1 , x_2 )$, Alice and Bob can use the
singlets $\Psi_\delta$ as a shared quantum information resource
to (i) synchronize their clocks if $\delta =0$ , and (ii) measure
and purify $\delta$ if they have synchronized clocks to start with.
In the general case of an unknown $\delta$ and an unknown time
synchronization offset, however, $\delta$ by itself is
not observable, and, consequently, $\Psi_\delta$ cannot
be purified without first establishing time synchrony between the two parties.

{\bf \noindent 4. Some promising directions for future research}

By using entangled (energetic) qubits as a resource shared between
spatially separated observers,
the QCS protocol as reformulated above allows the direct measurement of
certain nonlocal, covariant phase functions on spacetime. Moreover,
this functionality of the protocol
is straightforward to generalize
to many-particle entanglement\cite{Ftnt3}. While these results give
first hints of a profound connection between quantum information and
spacetime structure, they fall just short of providing a practical clock
synchronization algorithm because of the uncontrolled phase offsets
[like $\delta$ in Eq.\,(7)] that arise inevitably during the
distribution of entanglement. Since, as we showed above, these phase
``errors" cannot be purified asynchronously after they are already in
place, a succesful completion of the (singlet based) QCS algorithm would need
some method of entanglement distribution which avoids the accumulation
of relative phase offsets. We believe a complete
clock synchronization algorithm
based on quantum information theory will likely result from one of the
following approaches:

{\it ``Phase-locked" entanglement distribution}: It may be possible to
use the singlet states' inherent
non-local (Bell) correlations (which remain untapped in the
current QCS protocol) to
implement a ``quantum feedback loop," which, during entanglement
transport, will help keep the phase offset $\delta$ vanishing to within a small
tolerance of error. For example, states of the form
\begin{equation}
\frac{1}{\sqrt{2}} \, ( \, |0\rangle_A |1\rangle_{A'} |1\rangle_B |0
\rangle_{B'} \, - 
\, |1\rangle_A |0\rangle_{A'} |0\rangle_B |1
\rangle_{B'} \, ) \; ,
\end{equation}
where two pairs of atoms (the primed and the unprimed pair) are
entangled together, are not only dark but also immune to phase offsets
during transport of the pairs to Alice and Bob (provided both pairs are
transported along a common worldline through the same external
environment). Can such ``phase-error-free" states be used to control the
purity of singlets during transport?

{\it Entanglement distribution without transport}: Physically moving
each prior-entangled subsystem to its separate spatial
location is not the only way to distribute entanglement. An intriguing idea
recently discussed by Cabrillo {\it et.\ al.} \cite{Zoller}
proposes preparing two spatially separated atoms in their long-lived excited
states $|1\rangle_A |1\rangle_B$. A single-photon detector that cannot even in
principle distinguish the direction from which a detected photon arrives
is placed half way between the atoms. When one of the atoms makes a transition
to its ground state and the detector registers the emitted photon,
the result of its measurement is to put the combined two-atom
system in the entangled state
\begin{equation}
\frac{1}{\sqrt{2}} \, ( \, |0 \rangle_A |1 \rangle_B + e^{i \phi}
|1\rangle_A |0\rangle_B \, ) \; ,
\end{equation}
where $\phi$ is a random phase. Is there a similar procedure (based on
quantum measurements rather than physical transport) which creates
entanglement with a controlled rather than random phase offset $\phi$?

{\it Avoiding entanglement distribution altogether}: Can classical
techniques of clock synchronization be improved in accuracy
and noise performance by combining them with techniques from quantum
information theory which do not necessarily involve (energetic) entanglement
distribution? A recent proposal in this direction was made by Chuang in
\cite{Chuang}.

\begin{acknowledgements}
We would like to acknowledge valuable discussions with Daniel Abrams,
George Hockney and Colin Williams of the
JPL Quantum Computing Technologies group.

The research described in this paper
was carried out at the Jet Propulsion Laboratory,
California Institute of Technology, under a contract with
the National Aeronautics and
Space Administration, and was supported by a contract
with the Department of Defense.
\end{acknowledgements}


\begin{references}
\bibitem{QCS} R.\ Jozsa, D.\ S.\ Abrams, J.\ P.\ Dowling and C.\ P.\ Williams,
Phys.\ Rev.\ Lett.\ {\bf 85}, 2010 (2000).
\bibitem{Ftnt1} There is one additional complication of the QCS
protocol, having to do with making sure that the Hadamard transforms Eqs.\,(3)
performed by Alice and Bob are phase matched. This matching is
equivalent to setting up a canonical isomorphism between the Hilbert
spaces $\SH_A$ and $\SH_B$, and
is relatively straightforward to implement as discussed in detail
in\cite{QCS}. Throughout
the rest of this letter, we will assume without further
comment that such an isomorphism has been set up by Alice and Bob
prior to the operations under discussion; \cite{QCS} should be
consulted for further details.
\bibitem{Telep} C.\ H.\ Bennett, G.\ Brassard, C.\ Cr\'{e}peau,
R.\ Jozsa, A.\ Peres,
and W.\ K.\ Wootters, Phys.\ Rev.\ Lett.\ {\bf 70}, 1895 (1993).
\bibitem{Crypt} A.\ K.\ Ekert, Phys.\ Rev.\ Lett.\ {\bf 67}, 661 (1991).
\bibitem{Bent3} C.\ H.\ Bennett, D.\ P.\ DiVincenzo, J.\ A.\ Smolin, and W.\ K.\
Wootters, Phys.\ Rev.\ {\bf A 54}, 3824 (1996).
\bibitem{Ftnt2} If the phase offset $\delta$ were
known, such an equivalent entangled-clock expression for $\Psi_\delta$
would be available in terms of modified clock states
$| {\pm}_\delta \rangle_A \equiv (|0\rangle_A \pm e^{i \delta /2} |1\rangle_A )
/ \sqrt{2}$, and $ |{\pm}_\delta \rangle_B \equiv
(|0\rangle_B \pm e^{-i \delta /2} |1\rangle_B )
/ \sqrt{2}$;
but if $\delta$ is known, Alice and Bob can instead simply purify $\Psi_\delta$
by agreeing to rotate Bob's $|0 \rangle$ basis state by $e^{i \delta}$.
\bibitem{Ftnt3} For example, to states of the form
\[
\Psi_N = \frac{1}{\sqrt{N ! }} \sum_{\sigma}
(-1)^{\sigma} |0 \rangle_{A \sigma (1)}
|1 \rangle_{A \sigma (2)}
\cdots |N-1 \rangle_{A \sigma (N)} \; ,
\]
where $A1, \; A2, \cdots, AN$ denote $N$ observers who have, distributed
to them, $N$ identical atoms with $N$ distinct internal energy levels,
and the sum $\sigma$ is over all permutations of $\{ 1,2,\cdots,N \}$.
The state $\Psi_N$ is a generalization of the singlet
$\Psi$ [Eq.\,(2)] in that it is invariant under $U \otimes
U \otimes \cdots \otimes U$ for arbitrary unitary $U$.
\bibitem{Zoller} C.\ Cabrillo, J.\ I.\ Cirac, P.\ Garcia-Fernandez,
and P.\ Zoller, Phys.\ Rev.\ {\bf A 59}, 1025 (1999).
\bibitem{Chuang} I.\ L.\ Chuang, Phys.\ Rev.\ Lett.\ {bf 85}, 2006
(2000).






\end{references}
\end{document}